\documentclass[article]{jss}

\usepackage{thumbpdf,lmodern}

\usepackage[]{algorithm2e}

\usepackage[utf8]{inputenc}

\usepackage{amsmath}

\shortcites{IshwaranCompetingRisks}
\shortcites{wihs}

\shortcites{survival_event_history_book}
\shortcites{IshwaranSurvival}
\shortcites{WolbersConcordanceCompetingRisks}

\author{Joel Therrien\\Simon Fraser University   \And Jiguo Cao\\Simon Fraser University}
\Plainauthor{Joel Therrien, Jiguo Cao}

\title{Random Competing Risks Forests for Large Data}

\Abstract{
Random forests are a sensible non-parametric model to predict competing risks data according to some covariates.  However, there are currently no packages that can adequately handle large datasets ($n > 100,000$). We introduce a new \proglang{R} package, \pkg{largeRCRF}, using the random competing risks forest theory developed by \cite{IshwaranCompetingRisks}. We verify our package's validity and accuracy through simulation studies, and show that its results are similar enough to \pkg{randomForestSRC} while taking less time to run. We also demonstrate the package on a large dataset that was previously inaccessible, using hardware requirements that are available to most researchers.
}

\Plainkeywords{competing risks, random forests, survival, R, Java}

\Address{
  Jiguo Cao\\
  Department of Statistics \& Actuarial Science\\
  Simon Fraser University\\
  Room SC K10545\\
  8888 University Drive\\
  Burnaby, B.C.\\
  Canada V5A 1S6\\
  E-mail: \email{jiguo.cao@sfu.ca}
}

\begin{document}




\section{Introduction} \label{sec:intro}

In a competing risks problem we are often concerned with finding the distribution of survival times \citep{survival_event_history_book, methods_for_lifetime_data_book} for subjects, when there are multiple mutually exclusive ways for a subject to terminate. In addition, some subjects' times are censored, in which case it's known that they survive at least up to a certain time. We also may have some covariates that we believe affect the distribution of survival times, and are interested in estimating the survival time distribution conditional on the covariates.

Traditional models for incorporating covariates, such as the semi-parametric Fine and Gray proportional sub-hazards model \citep{FineAndGrayProportional}, often impose some parametric assumptions on the covariates which are useful when data is sparse but restrictive when data is plentiful. Random competing risks forests \citep{IshwaranCompetingRisks} are a non-parametric model based on Breinam's random forest algorithm \citep{Breiman2001} for where the response is competing risks data. A popular package \pkg{randomForestSRC} \citep{IshwaranRfsrc,IshwaranSurvivalR,IshwaranSurvival} is developed as an \proglang{R} package to train random competing risks forests, which works well for small and medium datasets. 

Unfortunately, \pkg{randomForestSRC} struggles to handle competing risks datasets greater than approximately 100,000 rows; with the required time to train a forest quickly growing to unfeasible levels. We introduce a new package, \pkg{largeRCRF}, that can train large datasets in a reasonable amount of time on consumer level hardware.

We will demonstrate using simulations that \pkg{largeRCRF} is able to produce similar enough accurate results to \pkg{randomForestSRC} while running many times faster (500x faster at $n=100,000$). We will also demonstrate how to use it on two real-life datasets; a small one to demonstrate using the package and a 1.1 million row dataset that was previously too large for this type of analysis.

\section{Theory of random competing risks forests} \label{sec:theory}

Random forests are a sequence of binary decision trees trained on bootstrap resamples of the original data. Assume we have some response and multiple predictor variables. For every tree (out of \code{ntree} trees), we bootstrap the training data and run \code{processNode(data, depth=0)} on it, which recursively grows the entire tree. A simplified version of \code{processNode} without edge cases is described below:

\renewcommand\AlCapFnt{\normalfont}

\begin{algorithm}
 \SetKwData{Data}{data}
 \SetKwData{Depth}{depth}
 \SetKwData{nodeSize}{nodeSize}
 \SetKwData{maxNodeDepth}{maxNodeDepth}
 \SetKwData{numberOfSplits}{numberOfSplits}
 \SetKwData{mtry}{mtry}
 \SetKwData{bestSplit}{bestSplit}
 \SetKwData{predictor}{predictor}
 \SetKwData{splitDataLeft}{splitDataLeft}
 \SetKwData{splitDataRight}{splitDataRight}
 \SetKwData{childNodeLeft}{childNodeLeft}
 \SetKwData{childNodeRight}{childNodeRight} 
 
 \SetKwFunction{processNode}{processNode}

 \SetKwProg{Fn}{Function}{}{}
 \Fn{\processNode{\Data, \Depth}}{
    \eIf{size of \Data $\geq  2 \times \nodeSize$ or \Depth < \maxNodeDepth}{
       \tcc{Split node}
       randomly select \mtry predictor variables\;
       \bestSplit $\leftarrow$ NULL\;
       \For{each selected \predictor}{
            \eIf{$\numberOfSplits = 0$}{
                Evaluate every possible split on \predictor, replacing \bestSplit with any better splits\;
            }{ 
                Evaluate \numberOfSplits random splits on \predictor, replacing \bestSplit with any better splits\;
            }
       }
       Using \bestSplit, produce \splitDataLeft and \splitDataRight\;
       \childNodeLeft $\leftarrow$ \processNode{\splitDataLeft, $\Depth+1$}\;
       \childNodeRight $\leftarrow$ \processNode{\splitDataRight, $\Depth+1$}\;
       return a split node with \bestSplit, \childNodeLeft, and \childNodeRight\;
       
    }{ 
        \tcc{Won't split; return a terminal node}
        return an average of \Data\;
    }
 }

 \caption{ How a node is created. A tree is created by running \code{processNode} on the root node's data, which recursively creates all of the child nodes under it.} \label{algorithm:processNode}

\end{algorithm}

When we make predictions using the forest, we follow the split nodes down to the appropriate terminal node on every tree and then average them together across the forest. There are three details that are not specified and vary based on the type of random forest used:
\begin{itemize} \setlength\itemsep{0em}
 \item How to calculate a score to find the best possible split.
 \item How to average the responses to form a terminal node.
 \item How to average across terminal nodes in the forest to make a prediction.
\end{itemize}

In order to train a random competing risks forest, we need to specify these details. We cover the theory that \cite{IshwaranCompetingRisks} developed, with minor differences highlighted. 

In terms of competing risks notation, let $T_i$ be the true termination time that happened for (or will happen for) subject $i$, and $\Delta_i$ be a status code for the type of event that ended / will end for subject $i$, $\Delta_i = 1, \dots, \text{J}$. Let $C_i$ be the censoring time that has / would have censored for subject $i$. Let $\tilde{T}_i = \min(T_i, C_i)$; it is the time that we actually have recorded in our dataset. Let $\delta_i = \Delta_i I(C_i < T_i)$; it is a status code for the type of event that ended for subject $i$, taking on a value of 0 if the subject was censored. 

Let us restrict ourselves to a node in a decision tree that is currently being trained, and suppose that there are $m$ observations in this node; the data we then work with is $\{ (\tilde{T}_i, \delta_i, X_i) | i=1, \dots, m \}$.

Definitions:
\begin{itemize}
 \item Let $Y(t) = \sum_{i=1}^m I(\tilde{T}_i \geq t)$, which gives the number of individuals at risk at time $t$.
 \item Define $d_j(t) = \sum_{i=1}^m I(\tilde{T}_i = t \cap \delta_i = j)$ which gives the number of events of type $j$ that occur at time $t$.
 \item For discussing splitting rules, suppose that a potential split produces a `left' and a `right' group. Denote $L$ and $R$ to be the set of observation indices of the parent node that would get assigned to each group. Define $Y_L(t)$, $Y_R(t)$, $d_{Lj}(t)$, $d_{Rj}(t)$ for each side as above, i.e.,
 \begin{itemize}
    \item $Y_L(t) = \sum_{i=1}^m I(\tilde{T}_i \geq t \cap i \in L)$; $Y_R(t) = \sum_{i=1}^m I(\tilde{T}_i \geq t \cap i \in R)$
    \item $d_{Lj}(t) = \sum_{i=1}^m I(\tilde{T}_i = t \cap \delta_i = j \cap i \in L)$; $d_{Rj}(t) = \sum_{i=1}^m I(\tilde{T}_i = t \cap \delta_i = j \cap i \in R)$
 \end{itemize}
 \item Index unique observed times $v_1, v_2, \dots v_K \in \{\tilde{T}_i | \delta_i \neq 0\}$ as an increasing sequence. 
\end{itemize}

\subsection{Splitting rules} \label{sec:theory:splitting_rules}

There are two choices for splitting rules. The first is the generalized log-rank test. For a fixed event type $j$, it corresponds to a test of the null hypothesis $H_0:\alpha_{Lj}(t) = \alpha_{Rj}(t) \quad \forall t \leq \max(\tilde{T}_i | \delta_i \neq 0)$, where $\alpha_{Lj}(t)$ and $\alpha_{Rj}(t)$ are the cause-specific hazard rates. It is calculated as follows:
\begin{equation}
    L_j^\text{LR} = \frac{1}{\hat{\sigma}_j^\text{LR}} \sum_{k=1}^{K} \bigg(d_{Lj}(v_k) - \frac{d_j(v_k) Y_L(v_k)}{Y(v_k)} \bigg)
    \label{eq:lr_split_score}
\end{equation} where
\begin{equation}
    (\hat\sigma_j^\text{LR})^2 = \sum_{k=1}^{K} \frac{Y_L(v_k)}{Y(v_k)} \bigg(1 - \frac{Y_L(v_k)}{Y(v_k)} \bigg) \bigg( \frac{Y(v_k) - d_j(v_k)}{Y(v_k) - 1} \bigg)
    \label{eq:lr_variance}
\end{equation}
When finding the best split, we try to find the split that maximizes $|L_j^\text{LR}|$. This test is restricted to only event $j$, but it can be calculated and combined for multiple events, where we will try to maximize $|L^\text{LR}|$ defined as
\begin{equation}
    L^\text{LR} = \frac{ \sum_{j=1}^\text{J} \hat\sigma_j^\text{LR} L_j^\text{LR} }{\sqrt{\sum_{j=1}^\text{J} (\hat\sigma_j^\text{LR})^2}}\,.
    \label{eq:lr_composite}
\end{equation} 
It should be noted that \cite{IshwaranCompetingRisks}  contains a typo (confirmed in correspondence with the first author), where they write Equation (\ref{eq:lr_composite}) with $(\hat\sigma_j^\text{LR})^2$ in the numerator instead of $\hat\sigma_j^\text{LR}$. This typo is only present in their paper; \pkg{randomForestSRC} correctly implements the splitting rule.

\cite{IshwaranCompetingRisks} also described a variant of this splitting rule that better handles competing risks data by instead calculating Gray's test \citep[Section 3.3.2]{IshwaranCompetingRisks}. This can be accomplished by reusing Equations (\ref{eq:lr_split_score}), (\ref{eq:lr_variance}), (\ref{eq:lr_composite}), while replacing $Y(t)$ with a cause-specific version when the censoring times are fully known.
\begin{equation}
    Y^*_j(t) = \sum_{i=1}^m I(\tilde{T}_i \geq t \cup (\tilde{T}_i < t \cap \delta_i \neq j \cap C_i > t))
    \label{eq:gray_risk_set}
\end{equation}
Note that \pkg{randomForestSRC} approximates (\ref{eq:gray_risk_set}) by using the largest observed time instead of the actual censor times (even if they're available), while \pkg{largeRCRF} explicitly requires that all censoring times be provided and does not yet support any approximate version.

\subsection{Creating terminal nodes} \label{sec:theory:terminal_node}

For generating a terminal node, we assume that the data has been split enough that it is approximately homogeneous enough to simply combine into estimates of the overall survival function, estimates of the cumulative incidence functions (CIFs), and estimates of the cumulative hazard functions using, respectively, the Kaplan-Meier estimator \citep{KaplanMeierCurve}, the Aalen and Johansen estimator \citep{AalenJohansenCIFs}, and the Nelson-Aalen estimator \citep{NelsonAalenEstimator1, NelsonAalenEstimator2}, which are expressed as %
\begin{eqnarray*}
    \hat{S}(t) = \hat{\Prob}(T \geq t) &=&\prod_{i=1}^{m(t)} \bigg(1 - \frac{\sum_{j=1}^J d_j(v_k)}{Y(v_k)} \bigg)\\
    \hat{F}_j(t) = \hat{\Prob}(T \geq t \cap \Delta = j) &=& \sum_{k=1}^{m(t)} \frac{\hat{S}(v_{k-1})d_j(v_k)}{Y(v_k)}\\
    \hat{H}_j(t) &=& \sum_{k=1}^{m(t)} \frac{d_j(v_k)}{Y(v_k)}
\end{eqnarray*}
where $m(t) = \max(k | v_k \leq t)$.

\subsection{Averaging terminal nodes} \label{sec:theory:averaging}

To make a prediction, we simply average the above functions across terminal nodes at each time $t$. To be specific, assume we have $M$ trees and are making a prediction for some predictors $X$. For each tree $k$ we follow the split nodes according to $X$ until we reach a terminal node, yielding functions $\hat{S}_k(t|X)$, $\hat{F}_{jk}(t|X)$, $\hat{H}_{jk}(t|X)$ for $\forall j=1 \dots J$. Then $\forall j=1 \dots J$ we define the overall functions that we return as:
\begin{eqnarray*}
    \hat{S}(t) &=& \frac{1}{M}\sum_{k=1}^M \hat{S}_k(t|X)\\
    \hat{F}_j(t) &=& \frac{1}{M}\sum_{k=1}^M \hat{F}_{jk}(t|X)\\
    \hat{H}_j(t) &=& \frac{1}{M}\sum_{k=1}^M \hat{H}_{jk}(t|X)
\end{eqnarray*}

\section{Simulation studies} \label{sec:simulation}

We run two simulation studies. The first simulation is used to verify the accuracy of \pkg{largeRCRF} by comparing it with \pkg{randomForestSRC} at a small sized dataset ($n=1000$). The second simulation is used to measure the time performance of both packages at varying simulation sizes. 

For the first simulation, we tune models for both packages on the data using the naive concordance error as described in \citet[Section 3.2]{WolbersConcordanceCompetingRisks} on a validation dataset, and then calculate an estimate of the integrated squared error on the CIFs using a final test dataset for the tuned models. We repeat this procedure 10 times for a sample size of $n=1000$. 

For the second simulation, we fix the tuning parameters and train both packages on simulated datasets of varying sizes, recording for each package the sum of the time used for training and the time used for making predictions on the validation dataset.

One important note; as of version 2.9.0 \pkg{randomForestSRC} adjusted their default algorithm to sample without replacement 63.2\% of the data for each tree, instead of using bootstrapping resampling normally associated with random forests. For these simulations we keep \pkg{randomForestSRC} at its previous default of bootstrap resampling.

We fix the number of trees to be trained at 100 for both simulations.

\subsection{Generating data}

Every training, validation, and test dataset in both simulations are generated in the same way. We first generate covariate vectors $X_1, X_2, X_3 \overset{iid}{\sim} N(0,1)$. We subset the space created by $X_1, X_2, \text{ and } X_3$ into 5 regions (see Table \ref{table:simulation_grid}). In each region, for each response to generate, we randomly select which competing risks event should occur based on prespecified probabilities, and then depending on the event, generate the competing risks time according to a prespecified distribution. By specifying the probabilities of each event and the distribution used to generate each event, we then know the true population cumulative incidence function (CIF) for any event $j$ (see (\ref{eq:cif_def})), which we can compare against the estimates.
\begin{equation}
\text{CIF}_j(t | X) = \Prob(T \leq t \cap \Delta=j | X)
           = \Prob(T \leq t | \Delta=j, X)\Prob(\Delta=j | X)
\label{eq:cif_def}
\end{equation}
Table \ref{table:simulation_grid} contains details on these weights and distributions. We also let censor times $C \sim \text{Exp}(\lambda={1}/{15})$, regardless of the covariates. We only allow \pkg{largeRCRF} and \pkg{randomForestSRC} access to $((\tilde{T}, \delta), X_1, X_2, X_3)$ where $\tilde{T}$ and $\delta$ are defined as in Section \ref{sec:theory}.

\begin{table}[t!]
\centering
    \begin{tabular}{p{0.45cm}|p{2cm}|p{2.1cm}|p{2.1cm}|p{3.5cm}|p{3.5cm}}
    \hline
    Set & Conditions & $\Prob(\Delta=1 | X)$ & $\Prob(\Delta=2 | X)$ & Dist. of $T | \Delta=1, X$ & Dist. of $T | \Delta=2, X$\\ \hline
    1 & $X_1 < 0$ \& $X_2 < 0$ \& $X_3 < 1$ & 0.4 & 0.6 & $\text{Weibull}(k=5, \lambda=6)$ & $\text{Exp}(1)$\\ \hline
    2 & $X_1 < 0$ \& $X_2 \geq 0$ \& $X_3 < 1$ & 0.1 & 0.9 & $\text{Lognormal}(0, 1)$ & Truncated positive $N(0,1)$\\ \hline
    3 & $X_1 \geq 0$ \& $X_2 < 0$ \& $X_3 < 1$ & 0.7 & 0.3 & $\text{Exp}(1)$ & $\text{Exp}(1)$ offset by $+1$\\ \hline
    4 & $X_1 \geq 0$ \& $X_2 \geq 0$ \& $X_3 < 1$ & 0.6 & 0.4 & $\text{Weibull}(k=1, \lambda=2)$ & $\text{Lognormal}(0, 1)$\\ \hline
    5 & $X_3 \geq 1$ & 0.5 & 0.5 & $\text{Exp}(\lambda=10)$ offset by $+2$ & $\text{Exp}(\lambda=\frac{1}{4})$ \\
    \hline
    \end{tabular}
    \caption{\label{table:simulation_grid} Overview of the distributions used to generate the simulated datasets.}
\end{table}

\subsection{The first simulation study - assessing accuracy} \label{sec:simulation:first}

\subsubsection{Tuning} \label{sec:simulation:first:tuning}

When we tune both packages for each of the 10 times, we want to maximize the concordance index error calculated on the validation dataset, except that we have quantities for each event to consider that aren't necessarily on the same scale. Let J be the number of events to consider (J=2). Let $\hat{C}_{ijk}$ be the concordance index as defined in \citet[Section 3.2]{WolbersConcordanceCompetingRisks} for tuning parameter combination $i$, event $j$, package $k$; evaluated with the predicted mortality of each validation observation being the integral of the estimated CIF for event $j$ for each of the observations from time 0 to the largest non-censored event time in the training dataset. $\hat{C}_{ijk}$ can be thought of as an estimate of the probability that the forest associated with $i$ and $k$ correctly predicts the ordering of two random event times for event $j$. 

We then calculate $\hat{C}_{ijk}^*$, the centered and scaled $\hat{C}_{ijk}$ according to $\underset{i}{\text{mean}}(\hat{C}_{ijk})$ and $\underset{i}{\text{sd}}(\hat{C}_{ijk})$, respectively, since we'd like to tune according to both events equally.

\begin{equation*}
 \hat{C}_{ijk}^* = \frac{\hat{C}_{ijk} - \underset{i}{\text{mean}}(\hat{C}_{ijk})}{\underset{i}{\text{sd}}(\hat{C}_{ijk})}
\end{equation*}

Then let the concordance error that we finally use to tune be
\begin{equation}
    \epsilon^\text{Concordance}_{ik} = - \frac{1}{\text{J}} \sum_{j=1}^\text{J} \hat{C}_{ijk}^*
    \label{eq:joint_concordance_error}
\end{equation}
We take the negative so that our intuition of minimizing error remains. For each package $k$, we then select the tuning parameters associated with the $i$ that minimized $\epsilon^\text{Concordance}_{ik}$, which we store for later use to calculate the CIF errors. 

The tuning parameters we consider are a grid formed by: 
\begin{itemize} \setlength\itemsep{0em}
    \item Number of splits tried (\code{nsplit}): [1, 50, 100, 250, 1000]
    \item Node size (\code{nodeSize}): [1, 10, 50, 100, 250, 500]
    \item Number of covariates tried at each split (\code{mtry}): [1, 3]
\end{itemize}

The splitting rule used for both packages is the composite log-rank rule defined in Equation (\ref{eq:lr_composite}) as it would not be equivalent to compare the two different Gray test implementations.

\subsubsection{CIF error} \label{sec:simulation:first:cif_error}

For the 10 selected models for each package we then calculate the error on the estimates of the cumulative incidence functions. For every observation $i$ and event $j$ in the test dataset we determine the true CIF according to Equation (\ref{eq:cif_def}) and Table \ref{table:simulation_grid}, which we denote as $\text{CIF}_{ij}$. Using the selected model trained on the training set we then calculate the corresponding estimate of the CIF for observation $i$ and event $j$ which we denote as $\widehat{\text{CIF}}_{ij}$.

Let $\tau=20$. We let $\tau$ be a constant number so that errors between training sets are comparable; 20 is otherwise arbitrary except that it encompasses most of the response times. We then calculate an error for each observation $i$ as follows:
\begin{equation*}
    \epsilon_{ij}^\text{CIF} = \sqrt{\int_0^\tau \bigr( \text{CIF}_{ij}(t) - \widehat{\text{CIF}}_{ij}(t) \bigr)^2\text{d}t}
    \label{eq:cif_error_individual}
\end{equation*}

We then average over the test dataset to calculate the mean error for event $j$.
\begin{equation*} 
    \epsilon_{\cdot j}^\text{CIF} = \frac{1}{n_\text{test}}\sum_{i=1}^{n_\text{test}} \epsilon_{ij}^\text{CIF}
    \label{eq:cif_error_event}
\end{equation*}

Finally we combine the CIF errors for our events together by averaging them.
\begin{equation*}
    \epsilon^\text{CIF} = \frac{1}{J}\sum_{j=1}^J \epsilon_{\cdot j}^\text{CIF}
\end{equation*}

\subsubsection{Putting it all together} \label{sec:simulation:first:putting_together}

We generate 10 training, validation, and test datasets with size 1000 each. Both \pkg{largeRCRF} and  \pkg{randomForestSRC} are trained on each training dataset with the different parameter combinations described earlier. $\epsilon^\text{Concordance}_{ik}$ is calculated on the validation dataset using each package's code for naive concordance  (as shown in Equation (\ref{eq:joint_concordance_error})). In addition, we also calculate $\epsilon^\text{Concordance}_{ik}$ for \pkg{randomForestSRC} using \pkg{largeRCRF}'s implementation of naive concordance (referred to as \pkg{randomForestSRC}/Alt), as there appears to be significant disagreement in the concordance error returned between the two packages.

For these three combinations (\pkg{largeRCRF}, \pkg{randomForestSRC}, and \pkg{randomForestSRC}/Alt), the sets of tuning parameters that minimized $\epsilon^\text{Concordance}_{ik}$ is calculated. Using the generated test sets, estimates of the error on the cumulative incidence functions are then calculated.

\subsubsection[Simulation results]{\textbf{Simulation results}} \label{sec:simulation:first:results}

\begin{figure}[t!]
\centering
\includegraphics{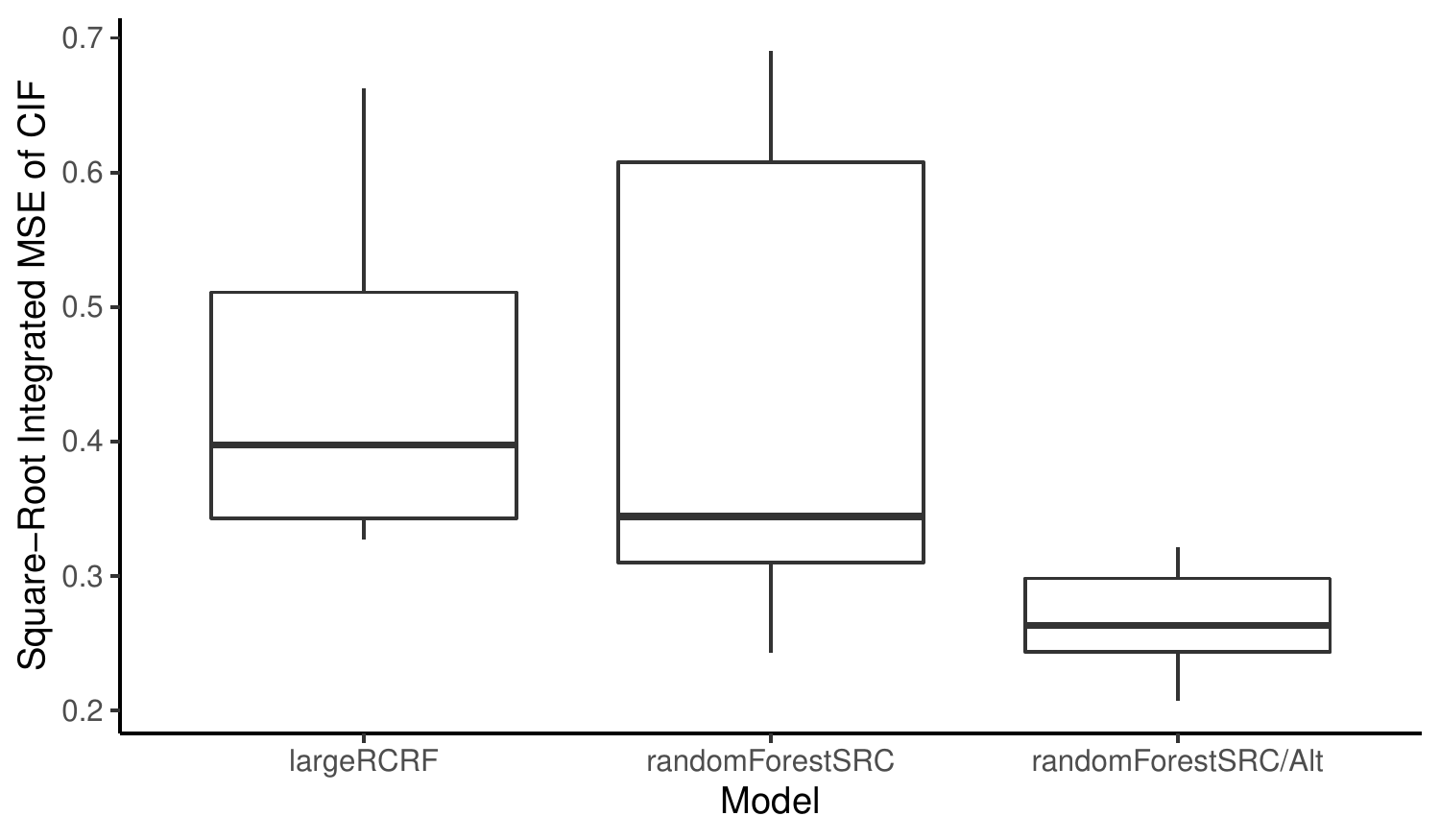}
\caption{\label{fig:simulation:raw_errors} Boxplot of cumulative incidence function (CIF) errors by model. \pkg{randomForestSRC}/Alt is the model produced by \pkg{randomForestSRC}, but tuned using the concordance error as calculated by \pkg{largeRCRF}.}
\end{figure}

Figure \ref{fig:simulation:raw_errors} shows the errors on the CIFs for the different models. 

\begin{figure}[t!]
\centering
\includegraphics{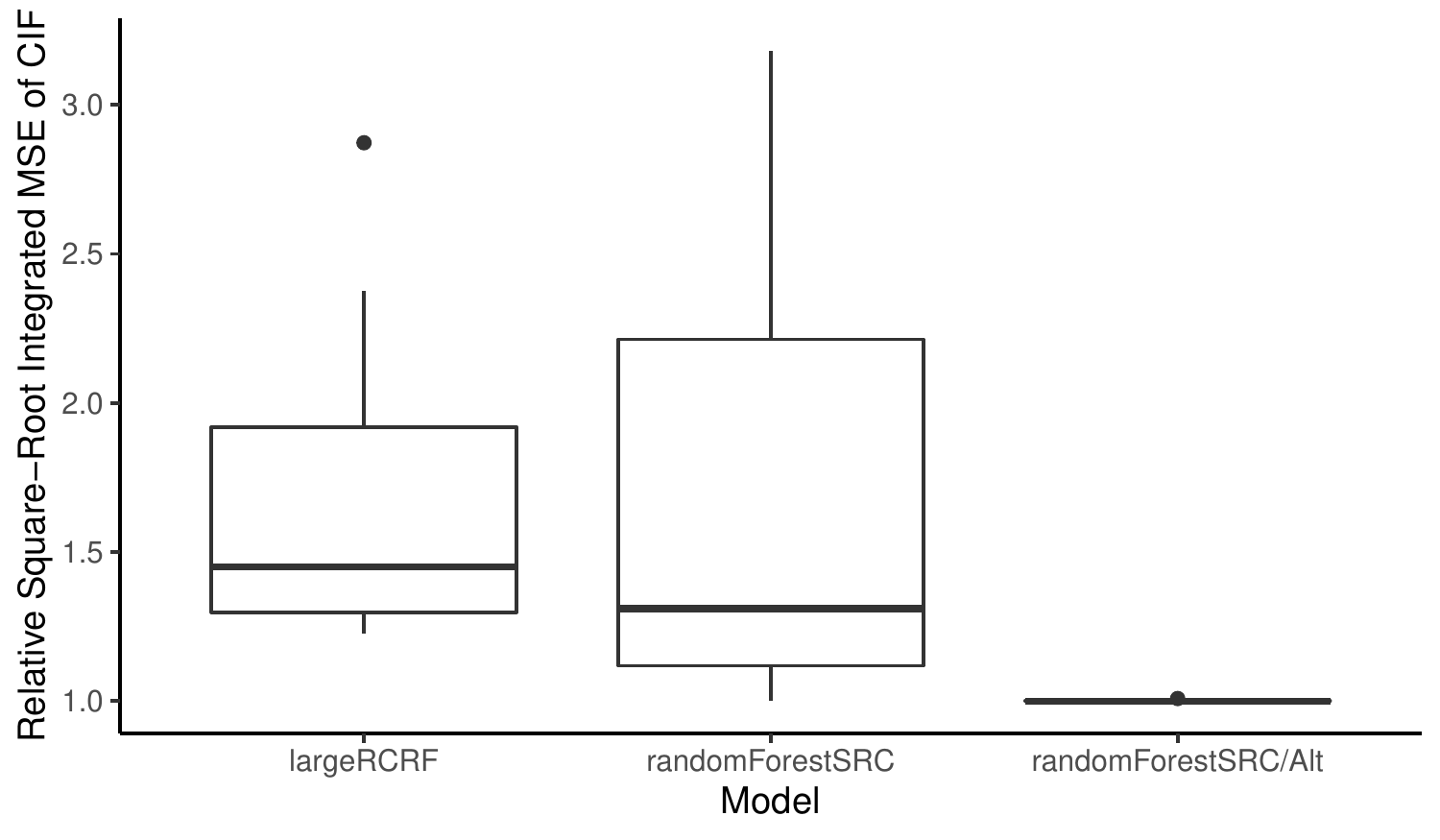}
\caption{\label{fig:simulation:relative_errors} Boxplot of relative cumulative incidence function (CIF) errors by model. For each dataset, all errors are divided by the smallest error produced; values of 1 means that that model produced the lowest error. \pkg{randomForestSRC}/Alt is the model produced by \pkg{randomForestSRC}, but tuned using the concordance error as calculated by \pkg{largeRCRF}.}
\end{figure}

Since all of the models consider the same dataset in each simulation, we can make the plot more informative by dividing each CIF error by the smallest error produced by the 3 models on that dataset. Figure \ref{fig:simulation:relative_errors} shows these results. 

It's clear from both plots that \pkg{largeRCRF} results are only slightly worse than \pkg{randomForestSRC}'s, demonstrating that \pkg{largeRCRF} is a viable alternative for researchers to use. Interestingly, \pkg{randomForestSRC} using \pkg{largeRCRF}'s implementation of naive concordance for tuning greatly outperforms both packages, suggesting that future versions of \pkg{randomForestSRC} may easily further improve their accuracy.

That said, the next simulation will show that \pkg{largeRCRF} runs significantly faster which could result in better performance if \pkg{largeRCRF} tuned on a denser grid.

\subsection{The second simulation - assessing speed} \label{sec:simulation:second}

For the timing simulation, we generate 6 datasets; two of size 1000, two of size 10,000, and two of size 100,000. For each pair of datasets we'll train forests on one and then use that forest to predict on the other, timing how long it takes to train and make predictions for each package, for each size dataset. For both packages we'll repeat the procedure 10 times for $n=1000$ and 5 times for $n=10,000$. For $n=100,000$ we train \pkg{largeRCRF} 5 times but \pkg{randomForestSRC} only once. The forest parameters used are constant and are:
\begin{itemize} \setlength\itemsep{0em}
    \item Number of splits tried (\code{nsplit}): 1000
    \item Node size (\code{nodeSize}): 500
    \item Number of covariates tried at each split (\code{mtry}): 1
\end{itemize}
\begin{table}[t!]
    \newcommand{\mc}[3]{\multicolumn{#1}{#2}{#3}}
    \centering
    \begin{tabular}{c|ccc|ccc|ccc}
     & \mc{3}{c|}{$n=1000$} & \mc{3}{c|}{$n=10,000$} & \mc{3}{c}{$n=100,000$}\\ \hline
    Package & Min. & Median & Max. & Min. & Median & Max. & Min. & Median & Max.\\ \hline
    \pkg{largeRCRF} & 2.10 & 2.17 & 2.32 & 35.7 & 36.2 & 37.0 & 736 & 742 & 748\\
    \pkg{randomForestSRC} & 7.04 & 7.14 & 7.21 & 3274 & 3350 & 3439 & - & 384061 & - \\ \hline
    \end{tabular}
    \caption{\label{table:timings} The recorded minimum, median, and maximum times (in seconds) recorded for training and predicting for each package for the different sizes of data. }
\end{table}
The simulations were run on a desktop computer running Linux with 16GB of RAM and an 8-core 3.5 GHz CPU.

The results, summarized in Table \ref{table:timings}, show that \pkg{largeRCRF} is faster than \pkg{randomForestSRC} at all three sizes. At $n=1000$ \pkg{largeRCRF} is only about 3x times faster, but that factor quickly grows as the sample sizes increases. At $n=10,000$ \pkg{largeRCRF} is about 90x faster, and at $n=100,000$ \pkg{largeRCRF} is about 500x faster. This speed increase allows researchers to perform more accurate and denser tuning, especially at larger sample sizes.

\section{Examples} \label{sec:example}

We'll next demonstrate \pkg{largeRCRF} on two real datasets. The first dataset will be a competing risks dataset from \textit{The Women's Interagency HIV Study} \citep{wihs} (retrieved from \pkg{randomForestSRC}). This dataset is relatively small at only 1164 rows and 4 possible predictors, making it a small and fast example to demonstrate how to use \pkg{largeRCRF}.

The second dataset is a much larger dataset from an online peer to peer lending company in the United States containing approximately 1.1 million rows and 76 possible predictors, which demonstrates training random competing risks forests on a dataset that was previously too large to work with.

\subsection{Women's Interagency HIV Study} \label{sec:example:wihs}

The \textit{Women's Interagency HIV Study} \citep{wihs} is a dataset that followed HIV positive women and recorded when one of three possible competing events occurred for each one:
\begin{itemize}\setlength\itemsep{0em}
 \item The woman began treatment for HIV.
 \item The woman developed AIDS or died.
 \item The woman was censored for administrative reasons.
\end{itemize}
There are four different predictors available (age, history of drug injections, race, and a blood count of a type of white blood cells).

The data is included in \pkg{largeRCRF}, but was originally obtained from \pkg{randomForestSRC}.
\begin{CodeChunk}
\begin{CodeInput}
R> data("wihs", package = "largeRCRF")
R> names(wihs)
\end{CodeInput}
\begin{CodeOutput}
[1] "time"     "status"   "ageatfda" "idu"      "black"    "cd4nadir"
\end{CodeOutput}
\end{CodeChunk}

\code{time} and \code{status} are two columns in \code{wihs} corresponding to the competing risks response, while \code{ageatfda}, \code{idu}, \code{black}, and \code{cd4nadir} are the different predictors we wish to train on. 

We specify \code{splitFinder = LogRankSplitFinder(1:2, 2)}, which indicates that we have event codes 1 to 2 to handle, but that we want to focus on optimizing splits for event 2 (which corresponds to when AIDS develops).

We specify that we want a forest of 100 trees (\code{ntree = 100}), that we want to try all possible splits when trying to split on a variable (\code{numberOfSplits = 0}), that we want to try splitting on two predictors at a time (\code{mtry = 2}), and that the terminal nodes should have an average size of at minimum 15 (\code{nodeSize = 15}; accomplished by not splitting any nodes with size less than 2 $\times$ \code{nodeSize}). \code{randomSeed = 15} specifies a seed so that the results are deterministic; note that \pkg{largeRCRF} generates random numbers separately from \proglang{R} and so is not affected by \code{set.seed()}.

\begin{figure}[t!]
\centering
\includegraphics{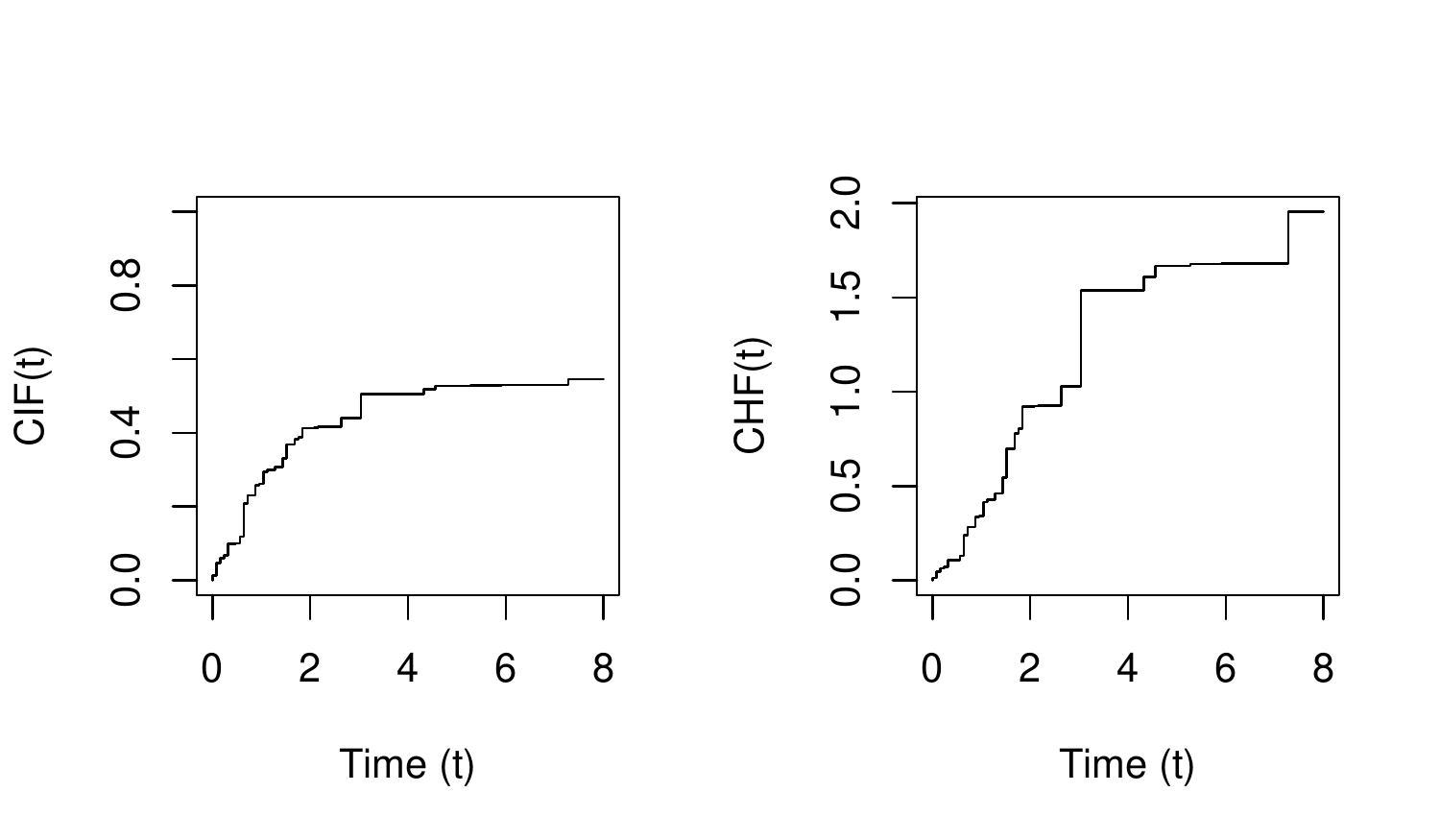}
\caption{\label{fig:example:wihs:plots} The out-of-bag estimated AIDS-specific cumulative incidence function (CIF) and cumulative hazard function (CHF) for subject 3.}
\end{figure}

\begin{CodeChunk}
\begin{CodeInput}
R> library("largeRCRF")
R> model <- 
+     train(CR_Response(status, time) ~ ageatfda + idu + black + cd4nadir,
+           data = wihs, splitFinder = LogRankSplitFinder(1:2, 2), 
+           ntree = 100, numberOfSplits = 0, mtry = 2, nodeSize = 15,
+           randomSeed = 15)
\end{CodeInput}
\end{CodeChunk}
Printing \code{model} on its own doesn't do much except print the different components and parameters that made the forest.
\begin{CodeChunk}
\begin{CodeInput}
R> model
\end{CodeInput}
\begin{CodeOutput}
Call:
train.formula(formula = CR_Response(status, time) ~ ageatfda + 
    idu + black + cd4nadir, data = wihs, splitFinder = LogRankSplitFinder(1:2, 
    2), ntree = 100, numberOfSplits = 0, mtry = 2, nodeSize = 15, 
    randomSeed = 15)

Parameters:
	Split Finder: LogRankSplitFinder(events = 1:2, eventsOfFocus = 2)
	Terminal Node Response Combiner: CR_ResponseCombiner(events = deltas)
	Forest Response Combiner: CR_FunctionCombiner(events = deltas)
	# of trees: 100
	# of Splits: 0
	# of Covariates to try: 2
	Node Size: 15
	Max Node Depth: 100000
Try using me with predict() or one of the relevant commands to determine error
\end{CodeOutput}
\end{CodeChunk}
We'll make predictions on the training data. Since we're using the training data, \pkg{largeRCRF} will by default only predict each observation using trees where that observation wasn't included in the bootstrap sample ('out-of-bag' predictions).
\begin{CodeChunk}
\begin{CodeInput}
R> predictions <- predict(model)
\end{CodeInput}
\end{CodeChunk}
Since our data is competing risks data, our responses are several functions which can't be printed on screen. Instead a message lets us know of several functions which can let us extract the estimate of the survivor curve, the cause-specific cumulative incidence functions, or the cause-specific cumulative hazard functions (CHF).
\begin{CodeChunk}
\begin{CodeInput}
R> predictions[[1]]
\end{CodeInput}
\begin{CodeOutput}
2 CIFs available
2 CHFs available
An overall survival curve available

See the help page ?CompetingRiskPredictions for a list of relevant functions
on how to use this object.
\end{CodeOutput}
\end{CodeChunk}
Here we extract the cause-specific functions for the AIDS event, as well as the overall survivor curve.
\begin{CodeChunk}
\begin{CodeInput}
R> aids.cifs = extractCIF(predictions, event = 2)
R> aids.chfs = extractCHF(predictions, event = 2)
R> survivor.curves = extractSurvivorCurve(predictions)
\end{CodeInput}
\end{CodeChunk}
Now we plot the functions that we extracted for subject 3; output in Figure \ref{fig:example:wihs:plots}.
\begin{CodeChunk}
\begin{CodeInput}
R> curve(aids.cifs[[3]](x), from=0, to=8, ylim=c(0,1),
+        type="S", ylab="CIF(t)", xlab="Time (t)")
        
R> curve(aids.chfs[[3]](x), from=0, to=8, 
+        type="S", ylab="CHF(t)", xlab="Time (t)")
\end{CodeInput}
\end{CodeChunk}

Finally, we calculate the naive concordance error on the out-of-bag predictions. \code{extractMortalities} calculates a measure of mortality by integrating the specified event's cumulative incidence function from 0 to \code{time}, although users are free to substitute their own measures if desired. \code{naiveConcordance} then takes the true responses and compares them with the mortality predictions we provide, estimating the proportion of wrong predictions for each event as described by \cite{WolbersConcordanceCompetingRisks}.

\begin{CodeChunk}
\begin{CodeInput}
R> mortalities1 <- extractMortalities(predictions, time = 8, event = 1)
R> mortalities2 <- extractMortalities(predictions, time = 8, event = 2)
R> naiveConcordance(CR_Response(wihs$status, wihs$time), 
+                list(mortalities1, mortalities2))
\end{CodeInput}
\begin{CodeOutput}
[1] 0.3939276 0.3535135
\end{CodeOutput}
\end{CodeChunk}

We could continue by trying another model to see if we could lower the concordance error, or by integrating the above steps into some tuning algorithm.

\subsection{Loan application} \label{sec:example:loans}

When a financial institution makes an installment loan, they may be interested in predicting when a potential borrower will prepay their loan, as this represents a loss of interest income for the financial institution. We have data of approximately 1.1 million three-year loans lent out by an online lender in the United States, containing 76 possible predictors, and the times of loan termination that represent one of:
\begin{itemize}
 \item How long the loan survived until the borrower defaulted on the loan.
 \item How long the loan survived until the borrower paid back the loan.
 \item The loan survived up to a time point (loan is censored due to when data was collected).
\end{itemize}
We also know for each loan which of the 3 events above occurred. In addition, since we know when the data was collected, we then know the censoring times for every loan, even if it had already terminated.

We would like to use this data to see if it is possible to accurately predict an 'expected loss of interest due to loan prepayment' that varies enough between borrowers that a financial institution could incorporate the information into their decision models when deciding what, if any, loan terms to offer a prospective borrower.

The response in the dataset includes only the time for when the loan terminated and not any dollar amounts of the prepayment. Thus we are restricted to a competing risks problem. With 76 predictors that have unknown effect on these times and over a million observations, assuming a parametric or semi-parametric model will make poor use of the data with bias dominating any estimates. Thus a non-parametric model like random forests should be considered.

Using \pkg{largeRCRF}, a forest of 100 trees was fit with the following parameters:
\begin{itemize} \setlength\itemsep{0em}
    \item Number of splits tried (\code{nsplit}): 1000
    \item Node size (\code{nodeSize}): 10,000
    \item Number of covariates tried at each split (\code{mtry}): 5
\end{itemize}   
We ignored maximum node depth.

This model was trained on a desktop computer with 16GB of RAM and an 8-core 3.5 GHz CPU, which is consumer-level hardware available to most researchers.

First we load the package. Since this is a large dataset we'll need to explicitly provide much of the system memory by setting the \proglang{Java} parameters. Note that \proglang{Java}'s garbage collector will also need memory, so we ask for a bit beneath 16GB.
\begin{CodeChunk}
\begin{CodeInput}
R> options(java.parameters = c("-Xmx14G", "-Xms14G"))
R> library("largeRCRF")
\end{CodeInput}
\end{CodeChunk}

Next we load the data. 
\begin{CodeChunk}
\begin{CodeInput}
R> trainingData <- read.csv("trainingData.csv.gz")
\end{CodeInput}
\end{CodeChunk}

Since our interest is only in the repayment event ($\delta=1$) and we have censor times available, we'll define a \code{GrayLogRankSplitFinder}. \pkg{largeRCRF} is designed to support different types of 'split finders' and 'response combiners', allowing other researchers to easily add support for other types of random forests using other types of responses without having to re-code the entire random forest algorithm.
\begin{CodeChunk}
\begin{CodeInput}
R> splitFinder <- GrayLogRankSplitFinder(events = 1:2, eventsOfFocus = 1)
\end{CodeInput}
\end{CodeChunk}

Due to how \proglang{Java} and \proglang{R} interface, at some point the data will unfortunately have to be duplicated from \proglang{R} into \proglang{Java}; which is problematic here because of the limited memory. \pkg{largeRCRF} supports deleting the dataset from \proglang{R}'s memory prior to beginning to train the forest, as long as the user provides an \proglang{R}'s environment object containing the data in place of a data frame.
\begin{CodeChunk}
\begin{CodeInput}
R> data.env <- new.env()
R> data.env$data <- trainingData
R> rm(trainingData)
\end{CodeInput}
\end{CodeChunk}

Finally we call \code{train}. We don't need to specify which response combiners we use; \pkg{largeRCRF} can infer that based on us using competing risks data. Since the memory on the computer used is limited and the dataset is large we can only afford to train one tree at a time, thus the \code{cores} parameter is explicitly set to 1. It should be noted that the other cores are not necessarily idle; \proglang{Java}'s garbage collector will require them, especially given the limited memory. The \code{savePath} parameter will have the package save trees to the hard drive as they are saved, which will save memory while training and allow us to recover our progress should training get interrupted.
\begin{CodeChunk}
\begin{CodeInput}
R> forest <- 
+     train(CR_Response(delta, u, censorTime) ~ ., data.env,
+     ntree = 100, numberOfSplits = 1000, mtry = 5, nodeSize = 10000,
+     cores = 1, splitFinder = splitFinder, savePath = "saved_forest")
\end{CodeInput}
\end{CodeChunk}

We want to train on all the available predictors, hence we used \code{.} on the right-hand side of the formula as is standard when using formulae in \proglang{R}. Training this model took about 10.5 hours.

Next we read in our test data and use it for predictions.
\begin{CodeChunk}
\begin{CodeInput}
R> testData <- read.csv("testData.csv.gz")
R> predictions <- predict(forest, testData)
\end{CodeInput}
\end{CodeChunk}

A list of \proglang{R} functions of the CIFs for repayment can then be extracted from the predictions.
\begin{CodeChunk}
\begin{CodeInput}
R> cif.predictions <- extractCIF(predictions, 1)
\end{CodeInput}
\end{CodeChunk}

\begin{figure}[t!]
\centering
\includegraphics{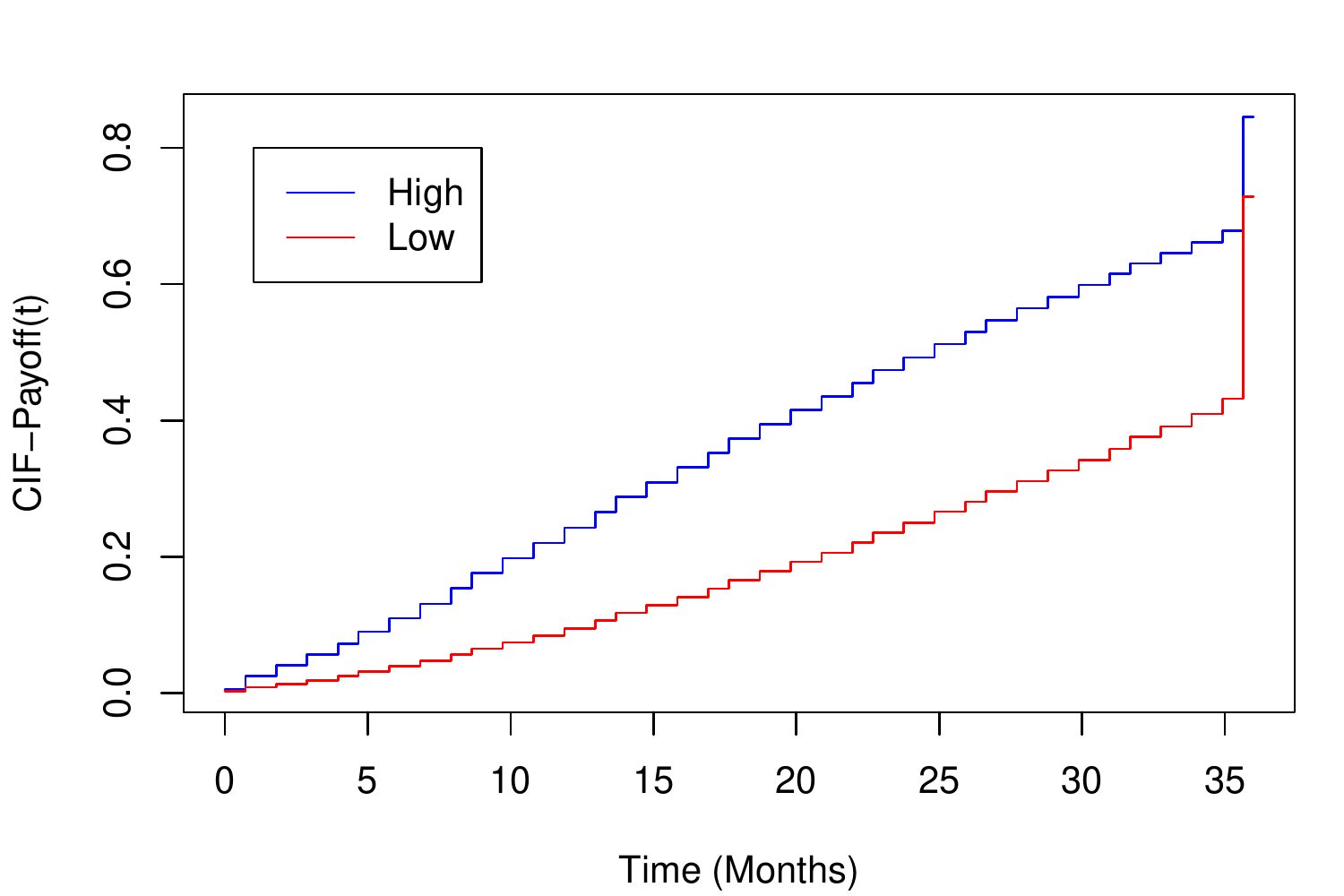}
\caption{\label{fig:cif_loans} Plot of the predicted repayment cumulative incidence functions (CIFs) for two loans with identical provided default-risk levels. The high jump at 36 months is associated with all of the loans that finished repaying right on time.}
\end{figure}

A pair of loans with vastly different repayment CIFs were selected such that they shared the same default-risk level as provided by the financial institution. The loans also were for the same loan amount (\$12,000), similar interest rates (6.97\%, 7.21\%), and similar monthly loan payments (\$370.37, \$371.68). We plot these curves together as shown in Figure \ref{fig:cif_loans}.
\begin{CodeChunk}
\begin{CodeInput}
R> pred_low <- cif.predictions[[20718]]
R> pred_high <- cif.predictions[[3073]]
R> curve(pred_high(x), from = 0, to = 36, type = 'S', col = 'blue',
+        ylab = "CIF-Payoff(t)", xlab = "Time (Months)")
R> curve(pred_low(x), from = 0, to = 36, type = "S", col = 'red',
+        add = TRUE)
R> legend(x = 1, y = 0.8, legend = c("High", "Low"), 
+        col = c("blue", "red"), lty = 1)
\end{CodeInput}
\end{CodeChunk}

We can now go back to the premise for this application: can we produce a measure of 'expected loss due to loan prepayment' that differs enough between borrowers that the financial institution thought were similar, such that the financial institution could use this new information in their decision models? Unfortunately we can only provide a range of values because the time at which a loan finishes being repaid does not exactly map to the quantity that a borrower prepaid each month. We'll instead assume two opposite extreme scenarios for how the borrower repaid to produce a range.
\begin{itemize} \setlength\itemsep{0em}
 \item \textbf{Prepayment Early}: The borrower immediately repaid a balance once they got the loan and then made regularly scheduled payments afterward, causing them to repay at the specified time. In this situation the financial institution lost the most because interest was not able to accrue on the immediately repaid balance.
 \item \textbf{Prepayment Late}: The borrower made regularly scheduled payments, and then on the day they repaid their loan they prepaid the entire remaining balance back. In this situation the financial institution lost the least because interest was able to accrue on the balance for the longest possible period.
\end{itemize}

For a given time $t$ (in months) for when the loan is repaid, we calculate the amount that the borrower would have paid in total under each of the two scenarios, and then calculate the difference between what the borrower would have paid in total had they never made any prepayments. This then gives us, under each scenario, the amount the financial institution lost due to prepayment for a given time $t$. We can define this as as functions $f(t)$ and $g(t)$, where $f$ is under the early scenario and $g$ is under the late scenario. Let $f(-1) = g(-1) = 0$ and say that a time of $-1$ denotes a loan that defaults (never finishes repaying); we thus let $f$ and $g$ record the amount lost due to prepayment only. As well, for time $t \geq 36$ let $f(t) = g(t) = 0$ as there was no prepayment.

Afterward we calculate the expected value for amount lost due to loan prepayment. Let $T$ be a discrete random variable for when a loan is repaid (units of months), taking on a value of $-1$ if the loan defaults. Then $\E[f(T)] = \sum_{t=-1}^{\infty} f(t) \cdot \Prob(T = t)$ where $\Prob(T = t)$ is extracted from the estimated CIF for loan repayment. $\E[g(T)]$ is calculated similarly.

For the two loans shown in Figure \ref{fig:cif_loans}, their expected losses due to prepayment are calculated and shown in Table \ref{table:expected_prepayments}. We see from the table that the difference in prepayment risk between the two loans under both assumptions are high. If a financial institution was able to somehow influence the behaviour of a borrower who is likely to repay early to instead repay later they could substantially increase their margins. 

\begin{table}
    \begin{centering}
    \begin{tabular}{p{2cm}|p{3cm}|p{3cm}}
    \textbf{Loan} & \textbf{Prepayments Early} & \textbf{Prepayments Late}\\ \hline
    High & $\$618$ & $\$318$\\
    Low & $\$342$ & $\$150$\\ 
    \hline \textbf{Difference} & $\$276$ & $\$169$
    \end{tabular}
    \caption{\label{table:expected_prepayments} Expected loss due to prepayment for two loans under the two assumptions of prepayment.}
    \end{centering}
\end{table}

While this example is somewhat contrived with the loans being artificially selected to maximize the difference in their CIFs, it does demonstrate that there is a potential gain for financial institutions to start looking into predicting loan prepayment behaviours. As well, this example demonstrates a random competing risks forest being trained on a dataset that was previously too large to work with, using only commonly available hardware.


\section{Summary} \label{sec:summary}

In this paper we implemented the random competing risks forest algorithm as described by \cite{IshwaranCompetingRisks} into \pkg{largeRCRF}, a software package designed to train random competing risks forests on large datasets. We ran simulations on smaller datasets to verify its accuracy as compared to an existing package, which showed that the package produces only slightly worse results. We also ran simulations on larger datasets which showed that for large datasets \pkg{largeRCRF} requires only a tiny fraction of the time that other packages require. We then demonstrated how to use \pkg{largeRCRF} using \textit{The Women's Interagency HIV Study}, a smaller dataset that most users readers can quickly follow. Finally, we demonstrated it on a large dataset of financial loan information that was previously too large for this type of analysis, finding that the required computational resources were reasonable for most researchers. This opens up random competing risks forests to more datasets and provides researchers an additional tool to try to explain their data.

\section*{When to use} \label{sec:when_to_use}

While \pkg{largeRCRF} trains random competing risks forests faster than \pkg{randomForestSRC}, it currently lacks many useful features such as calculating variable importance or missing data imputation. In general, unless researchers are working with a dataset that is too large for \pkg{randomForestSRC} they'll likely have an easier time with \pkg{randomForestSRC}.


\section*{Computational details}

Most of \pkg{largeRCRF} is written in \proglang{Java} 1.8, with an \proglang{R} interface available written using the \pkg{rJava} 0.9-10 package \citep{rJava}. There is also support for directly executing the \proglang{Java} code from a command-line interface.


The results in this paper were obtained using \proglang{R}~3.5.2 \citep{RCitation} with the \pkg{randomForestSRC}~2.9.0 package. Many visualizations were created using the \pkg{ggplot2}~3.1.1 package \citep{ggplot2}. \proglang{R} itself and all third-party packages used are available from the Comprehensive \proglang{R} Archive Network (CRAN) at
\url{https://CRAN.R-project.org/}. \pkg{largeRCRF} is available at \url{https://github.com/jatherrien/largeRCRF}.

\section*{Known issues}

Due to limitations in how \proglang{R} and \proglang{Java} interface, \proglang{R} at most only has references to different results produced by \proglang{Java} unless they were explicitly copied into \proglang{R}. This has two effects - 
\begin{itemize}
 \item \proglang{Java}'s garbage collector does not know whether \proglang{R} continues to have a reference to a \proglang{Java} object or not; therefore it never cleans up any objects given to \proglang{R}. This results in huge memory inefficiencies when \pkg{largeRCRF} is used over and over again in the same \proglang{R} session, as each previously trained forest cannot be removed from memory even if the user ran \code{rm()} on the object. A workaround is to restart the \proglang{R} session when you want to reclaim memory; a well-tested way to achieve this is to make use of the \pkg{parallel} \citep{RCitation} package and train forests in clusters of size 1, as this will train the forests in independent \proglang{R} sessions.
 \item When an \proglang{R} workspace is saved, it will save the references to the \proglang{Java} objects but not the \proglang{Java} objects themselves; meaning that the \proglang{Java} objects cannot be saved through the traditional \proglang{R} methods. The workaround for this is to use the \code{saveForest} and \code{loadForest} methods provided in \pkg{largeRCRF}.
\end{itemize}

It's been observed when running on a particular academic cluster system that sometimes \pkg{largeRCRF} stalls immediately before training a single tree, while on other systems it always run fine. In addition, using \proglang{Java}'s \code{parallelStream} on this particular system for making predictions has caused crashes. These issues has only been observed on this one particular system and is suspected to be an issue with its installed \proglang{Java} runtime; however that hasn't been confirmed. 

\section*{Acknowledgments}

Thank you to Dr. Hemant Ishwaran for answering my many questions and for developing the initial random competing risks forest algorithm. Thank you to Daniel Daly-Grafstein for helping go over my code to find bugs.


\bibliography{refs}


\newpage

\begin{appendix}

\section{Implementation Differences} \label{appendix:differences}

There are some implementation differences between \pkg{largeRCRF} and \pkg{randomForestSRC}. 

When Gray's test  is used as a splitting rule \citep[Section 3.3.2]{IshwaranCompetingRisks}, \pkg{randomForestSRC} approximates it by replacing the censor times used with the largest observed value. In \pkg{largeRCRF} direct censor times are used and must be included in the data as part of the response. Future versions of \pkg{largeRCRF} may support the approximate version as well.

There is also a difference in how the two packages handle missing values among the predictors. \pkg{randomForestSRC} trains an unsupervised forest to predict missing covariate values, which are then used to replace any missing values in the actual competing risks random forest. \pkg{largeRCRF} instead randomly assigns missing values after each split, as described in \citet[Section 2.3, steps 1-3]{FeiIshwaranMissingData}

As of version 2.9.0 \pkg{randomForestSRC} now by default samples 63.2\% of the data without replacement for each tree. \pkg{largeRCRF} instead performs standard bootstrap resampling, which was the previous default.
\end{appendix}


\end{document}